\documentclass[onecolumn,floatfix,superscriptaddress,a4paper,
               showpacs,showkeys,nofootinbib,preprint]{revtex4}
\usepackage{epsfig}
\usepackage{latexsym}
\usepackage{xspace}
\usepackage{hyperref}
\usepackage[latin2]{inputenc}
\usepackage{indentfirst}
\usepackage{enumerate}

\usepackage{amsmath}
\usepackage{amssymb}
\usepackage[english]{babel}
\usepackage{url}

\newcommand{\eq}[1]{\begin{align} #1 \end{align}}

\begin{document}

\title{Strongly Intensive Measures for
\\  Transverse Momentum and Particle Number Fluctuations
       }

 \author{Mark I. Gorenstein}
 \affiliation{Bogolyubov Institute for Theoretical Physics, Kiev, Ukraine}
 \affiliation{Frankfurt Institute for Advanced Studies, Frankfurt, Germany}
\author{Katarzyna Grebieszkow}
\affiliation{Faculty of Physics, Warsaw University of Technology, Warsaw, Poland}
\begin{abstract}
Strongly intensive measures $\Delta[P_T,N]$ and
$\Sigma[P_T,N]$ are used to study the event-by-event fluctuations
of the transverse momentum $P_T$ and particle multiplicity $N$ in
nucleus-nucleus collisions.
A special normalization for these fluctuation measures ensures
that they are dimensionless and yields a common scale required for
a quantitative comparison of fluctuations. In this paper basic
properties of the $\Delta[P_T, N]$ and $\Sigma[P_T, N]$ measures
are tested within different phenomenological models using the
Monte Carlo simulations (the so-called fast generators) and
analytical solutions. The obtained results are helpful to
elucidate the properties of the $\Delta[P_T,N]$ and $\Sigma[P_T,N]$
measures.

\end{abstract}



\maketitle
\section{Introduction}
The main motivation for the experiments studies on relativistic
nucleus-nucleus (A+A) collisions is to create and study
strongly interacting matter. Experimental and theoretical
investigations of event-by-event (e-by-e) fluctuations in A+A
collisions produce new information about its properties.
E-by-e fluctuations can be also an important tool for localizing the
phase boundary and the critical point of the QCD matter. In
particular, significant increase of transverse momentum and
multiplicity fluctuations are expected in a vicinity of the
critical point \cite{crit}. One can probe different regions of the
phase diagram by varying the collision energy and the size of
colliding nuclei \cite{size}. A possibility to observe signatures
of the critical point inspired the energy and system size scan
program of the NA61/SHINE Collaboration at the CERN Super Proton
Synchrotron (SPS) \cite{Ga:2009} and the low energy scan program
of the STAR and PHENIX Collaborations at the Brookhaven National
Laboratory's Relativistic Heavy Ion Collider  (BNL RHIC)
\cite{RHIC-SCAN}. In these studies one measures and then compares
the e-by-e fluctuations in collisions of different nuclei at
different collision energies. The average sizes of created
physical systems and their e-by-e fluctuations are expected to be
rather different \cite{KGBG:2010}. This strongly affects the
observed hadron fluctuations, and consequently
the measured quantities do
not describe local physical properties of the system but
rather reflect the system size fluctuations. For instance, A+A
collisions with different centralities may produce a system with
approximately the same local properties (e.g., the same
temperature and baryonic chemical potential) but with the volume
changing significantly from interaction to interaction. Note that
in high energy collisions the average volume  of created matter
and its variations from collision to collision are usually out of
experimental control, i.e. the volume variations are difficult or
even impossible to measure. Therefore, a suitable choice of
statistical tools for the study of e-by-e fluctuations is really
important.

In statistical mechanics, an extensive quantity is proportional to
the system volume $V$, whereas  an intensive one has  fixed finite
value in the thermodynamical limit $V\rightarrow \infty$.
Intensive quantities are used to describe local properties of a
physical system. In particular, the equation of state of the matter
is usually formulated in terms of intensive physical
quantities, e.g., the pressure is considered as a function of
temperature and chemical potentials.

The strongly intensive quantities have been introduced in
Ref.~\cite{GG:2011}. Within the grand canonical ensemble
formulation of statistical mechanics they are independent of the
average volume and volume fluctuations. Similar properties take
place in the model of independent sources: the strongly intensive
measures of fluctuations are independent of the average number of
sources and of fluctuations of the number of sources. The strongly
intensive measures $\Delta[A,B]$ and $\Sigma[A,B]$ are suggested
for studies of e-by-e fluctuations of hadron production in heavy ion
collisions at high energies.
They are defined using two arbitrary extensive quantities $A$ and
$B$. In the present paper we consider a pair of extensive
variables: the transverse momentum $A=P_T=p_T^{(1)}+\dots
p_T^{(N)}$, where $p^{(i)}_T$ is the absolute value of the
$i^{{\rm th}}$ particle transverse momentum, and the number of
particles $B=N$.
The measures $\Delta[P_T,N]$ and $\Sigma[P_T,N]$ were
studied recently within the ultra-relativistic quantum molecular dynamics
(UrQMD) simulations in Ref.~\cite{KG:APP2012}. The measures
$\Delta[A,B]$ and $\Sigma[A,B]$ in the case of two hadron
multiplicities $A$ and $B$ were considered within the
hadron-string dynamics (HSD) transport model in Ref.~\cite{HSD}.
To simplify notations we sometimes use $X=P_T$ and
$x_i=p_T^{(i)}$. Note that our consideration is valid also for
other motional variables $X$, e.g., the system energy
$X=E=\epsilon_1+\dots+\epsilon_N$.
The strongly intensive measure $\Delta[X,N]$ and $\Sigma[X,N]$
are defined as \cite{GG:2011}:
 \eq{\label{Delta-XN}
 &\Delta[X,N]
 ~=~ \frac{1}{C_{\Delta}} \Big[ \langle N\rangle\,
      \omega[X] ~-~\langle X\rangle\, \omega[N] \Big]~, \\
&\Sigma[X,N]
 ~=~ \frac{1}{C_{\Sigma}} \Big[ \langle N\rangle\,
      \omega[X] ~+~\langle X\rangle\, \omega[N] ~-~2\Big(\langle X\,N\rangle
      ~-~\langle X\rangle \langle N\rangle\Big)\Big]~,\label{Sigma-XN}
}
where
\eq{\label{omega-A}
\omega[X]~=~\frac{\langle X^2 \rangle~ -~ \langle X \rangle^2}{
\langle X \rangle}~,~~~~\omega[N]~=~\frac{\langle N^2 \rangle~ -~ \langle N \rangle^2}{
\langle N \rangle}~
}
are the scaled variances for $X$ and $N$ fluctuations, and
$C_{\Delta}$ and $C_{\Sigma}$ are normalization factors.
The notation $\langle \dots \rangle$ represents the e-by-e averaging.

The first strongly intensive measure of fluctuations, the
so-called  $\Phi$ measure, was introduced a long time ago in
Ref.~\cite{GM:1992}. There were many attempts  to use the $\Phi$
measure in the data analysis~\cite{Phi_data,d1,d2,d3,d4,d5,d6} and
in theoretical
models~\cite{Phi_models,m2,m3,m4,m5,m6,m7,m8a,m8,m9,m10,m11,MRW:2004,m12,m13,m14}.
In general, $\Phi$ is a dimensional quantity and it does not
assume a characteristic scale for a quantitative analysis of
e-by-e fluctuations for different observables. Note that the
latter properties  were clearly disturbing.

In the recent paper \cite{GGP:2013} special normalization has been
proposed for the $\Delta$ and $\Sigma$ fluctuation measures. It is
used in the present study and ensures that measures
(\ref{Delta-XN}) and (\ref{Sigma-XN}) are dimensionless and yields
a common scale required for a quantitative comparison of the
e-by-e fluctuations. This normalization has been already used for
the $\Delta[P_T,N]$ and $\Sigma[P_T,N]$ measures using the transport model
of A+A collisions in Ref.~\cite{GGP:2013} and for
the ideal quantum gases within the grand canonical ensemble
formulation \cite{GR:2013}. Note that the NA61
Collaboration has already started to use the strongly intensive measures
to study e-by-e fluctuations in A+A collisions \cite{NA61-plan}.

In the present paper several phenomenological models of hadron
production are suggested and studied using the Monte Carlo (MC)
simulations (the so-called fast generators). Analytical
solutions for the proposed models are also presented and analyzed.
These studies are helpful to elucidate  properties of the
$\Delta[P_T,N]$ and $\Sigma[P_T,N]$ measures. A search for
possible signals for the phase transition and critical point
in A+A collisions is outside of the scope of our paper.
To achieve this goal one first needs to formulate suitable dynamical
models for these phenomena.

The paper is organized as follows. In Sec.~\ref{ref-mod} we
introduce two reference models. The first model is the independent
particle model within which we calculate the normalization factors
$C_{\Delta}$ and $C_{\Sigma}$ for $\Delta[X,N]$ and $\Sigma[X,N]$
quantities. The second model is the model of independent sources
which is often used to analyze the data on nucleus-nucleus
collisions. Section~\ref{fast} presents examples of the MC
simulations. Some of these examples correspond to different
versions of the model of independent sources. Analytical
solutions are also presented and analyzed. In Sec.~\ref{TN-corr}
the MC simulations and analytical consideration are used for the
models where single particle momentum spectra are dependent on the
number of the produced particles. In Sec.~\ref{urqmd} results of statistical
and transport models are presented. Using the UrQMD simulations
we study effects of the centrality selection and limited detector acceptance and efficiency
in A+A collisions.
A summary in Sec.~\ref{sum} closes the article.

\section{Reference Models}\label{ref-mod}
In this section two simple models of particle production are
presented. The first one is the independent particle model (IPM)
which is used as a reference model to fix the normalization of the
strongly intensive measures $\Delta$ and $\Sigma$. Namely,
properly normalized strongly intensive quantities assume the value
{\it one} for the fluctuations given by the  IPM. The second model
is the model of independent sources. In this model, the values of
$\Delta$ and $\Sigma$ for the system of sources are equal to
their values for a single source.

\subsection{Independent Particle Model}
In Ref.~\cite{GGP:2013} a special normalization for the strongly
intensive measures $\Delta[A,B]$ and $\Sigma[A,B]$ has been proposed.
In this subsection we present its derivation when $A$ is an 
extensive variable $A=X$ presented as a sum of single particle terms
\eq{\label{X}
X~=~x_1~+x_2~+\ldots~+ x_{N}~,
}
(e.g., the system energy $E$ or transverse
momentum $P_T$)
and $B=N$ is the number of particles. Inter-particle
correlations are absent in the IPM, i.e. the probability of any
multi-particle state is a product of probability distributions
$F(x_j)$ of single-particle variables $x_j$, and these probability
distributions are the same for all $j=1,\ldots, N$ and independent
of the number of particles $N$~:
\eq{ \label{dist-X}
F_N (x_1,x_2, \dots, x_N)~ = ~{\cal P}(N)\times
F(x_1)\,F(x_2)\times \cdots \times F(x_N)~,
}
where ${\cal P}(N)$ is an arbitrary multiplicity distribution of
particles. The functions entering Eq.~(\ref{dist-X}) satisfy the
normalization conditions:
\eq{\label{norm-P}
\sum_N {\cal P}(N)~=~1~,~~~~~\int dx~F(x)~=~1~.
}
The averaging procedure for $k^{{\rm th}}$ moments of any
multiparticle observable $A$ reads:
\eq{\label{av}
\langle A^k \rangle ~=~\sum_{N}{\cal P}(N) \int dx_1dx_2\ldots dx_N  F(x_1)\,F(x_2)\times
\cdots \times F(x_N)~\Big[A(x_1,x_2,\ldots,x_N)\Big]^k~.
}
For the first and second moments of $X$ and $N$ one obtains:
\eq{\label{IPM-X}
& \langle X\rangle =
\overline{x}\cdot \langle N\rangle~,~~~~
%
\langle X^2\rangle =\overline{x^2}\cdot\langle N\rangle
+\overline{x}^2\cdot\left[\langle N^2\rangle -\langle
N\rangle\right]~,~~~~
%
\langle XN\rangle =\overline{x}\cdot \langle N^2\rangle~,
%
 }
where
\eq{
\langle N^k\rangle ~=~\sum_N {\cal P}(N)\,
N^k~,~~~~~\overline{x^k}~=~\int dx\, F(x)\, x^k~.
}
Note that the overline denotes averaging over single particle
inclusive distribution, whereas $\langle \dots \rangle$ represents
event averaging over multiparticle states of the system,
e.g., e-by-e averaging over hadrons detected in A+A collisions.

Using Eq.~(\ref{IPM-X}), one finds
\eq{\label{omega-X-IPM}
& \omega[X]\equiv~\frac{\langle X^2\rangle -\langle
X^2\rangle}{\langle
X\rangle}~=~\frac{\overline{x^2}-\overline{x}^2}{\overline{x}}~+~
\overline{x}\cdot \frac{\langle N^2\rangle -\langle N\rangle^2}{\langle N\rangle}
~\equiv~\omega[x]~+~\overline{x}\cdot \omega[N]~,\\
&\langle XN\rangle ~-~\langle X\rangle \,\langle N\rangle
~=~\overline{x}\cdot \Big[\langle N^2\rangle ~-~\langle
N\rangle^2\Big]~\equiv~\overline{x}\cdot \langle N\rangle~ \omega[N]~,\label{XN-IPM}
}
 and finally,
 \eq{\label{Delta-IPM}
 &\Delta[X,N]
 ~=~ \frac{1}{C_{\Delta}}~ \Big[ \langle N\rangle\,
      \omega[X] ~-~\langle X\rangle\, \omega[N] \Big]~
      =~\frac{\omega[x]\cdot \langle N\rangle}{C_{\Delta}}~, \\
&\Sigma[X,N]
 ~=~ \frac{1}{C_{\Sigma}} ~ \Big[ \langle N\rangle\,
      \omega[X] ~+~\langle X\rangle\, \omega[N] ~-~2\Big(\langle X\,N\rangle
      ~-~\langle X\rangle \langle N\rangle\Big)\Big]~
      =~\frac{\omega[x]\cdot \langle N\rangle}
      {C_{\Sigma}}~ .\label{Sigma-IPM}
}
The requirement that
\eq{\Delta[X,N]~=~\Sigma[X,N]~=~1 \label{DelSig=1}
}
for the  IPM leads thus to the normalization factors
%
%
%
\eq{\label{norm}
 C_{\Delta}~=~C_{\Sigma}=~
\omega[x]\cdot \langle N\rangle~,~~~~~
 \omega[x]~\equiv~\frac{\overline{x^2}~-~\overline{x}^2}
{\overline{x}}~.
}
The normalization factors (\ref{norm}) are suggested to be used both
in theoretical models and for the data analysis (see
Ref.~\cite{GGP:2013} for further details of the normalization
procedure).

According to the current classification the $\Phi$ measure
\cite{GM:1992} belongs to the $\Sigma$ family~\cite{GG:2011}. It
can be calculated as
\eq{\label{Phi}
\Phi_X~=~\Big[\overline{x}\,\omega[x]\Big]^{1/2}\,\Big[\sqrt{\Sigma[X,N]}
~-~1\Big]~.
}

The representation of $X$ with Eq.~(\ref{X}) as the sum of single
particle variables $x_i$ is an evident feature of the  IPM. Thus,
one needs such a representation to calculate the normalization
factors $C_{\Delta}$ and $C_{\Sigma}$. Such a representation of
the extensive motional variable is, however, not necessarily needed
for the e-by-e measurements. For example, the system energy $E$ (or
transverse momentum $P_T$) can be measured by a calorimeter
without determining individual single particle contributions.

It was proven \cite{GGP:2013} that the IPM relation (\ref{DelSig=1})
is valid also in two models. The first model is statistical mechanics for
the Boltzmann ideal gas within the grand canonical ensemble.
The second model is the {\it mixed event procedure} which creates a sample of
artificial events, where each particle is taken from different physical events.
These model constructions play an important role as  {\it reference
model}. The deviations of real data from the IPM
results (\ref{DelSig=1}) can be used to clarify the physical
properties of the system. It resembles the situation with particle
number distributions. One prefers to use the Poisson distribution
%
%
$P(N)=\exp\left(-\,\overline{N}\right)\,\overline{N}^N/{N!}$~
%
%
with $\omega[N]=1$ as a reference model. Another reference value
$\omega[N]=0$ corresponds to $N={\rm const}$, where the $N$-fluctuations
are absent. The fluctuations for
any particle number distribution ${\cal P}(N)$ is then clarified
by the comparison of the calculated (or measured) scaled variance
$\omega[N]$ with its reference value of $\omega[N]=1$.
The relation $\omega[N]>1$ (or $\omega[N]\gg 1$) corresponds to ``large''
(or ``very large'') fluctuations of $N$, and $\omega[N]<1$
(or $\omega[N]\ll 1$) to ``small''
(or ``very small'') fluctuations.

\subsection{Model of Independent Sources}
In this subsection we consider a model of independent sources
(MIS) for multi-particle production.   In this model the number of
sources, $N_S$, changes from event to event. The sources are
statistically identical and independent of each other. A famous
example of  the MIS is the wounded nucleon model \cite{WNM} for
A+A collisions . Two fluctuating extensive quantities $X$ and $N$
can be expressed as
\eq{\label{ind-sour}
X~=~X_1~+X_2~+\ldots~+ X_{N_S}~,~~~~~N~=~n_1~+n_2~+\ldots~+
n_{N_S}~,
}
where   $n_j$ denotes the number of particles emitted from the
$j^{{\rm th}}$ source ($j=1,\ldots,N_S$), and
$X_j=x_1+\dots+x_{n_j}$ is the contribution from the $j^{{\rm
th}}$ source to the quantity $X$.

Overline notations will be used for the averages connected to a
single source. The single-source quantities are independent of
$N_S$ and have the properties of intensive quantities. The
single-source distribution $F_S(X_S,n)$ is assumed to be
statistically identical for all sources, thus, for all
$j=1,\dots,N_S$ it follows:
\eq{\label{id-source}
\overline{X_j^k}~\equiv ~\overline{X_S^k}~,~~~~
\overline{n_j^k}~\equiv ~\overline{n^k}~,~~~
\overline{X_jn_j}~\equiv ~\overline{X_Sn}~,
}
where  $\overline{X_S^k}$, $\overline{n^k}$, and
$\overline{X_S\,n}$  (for $k=1,2$) are the first and second
moments of the distribution $F_S(X_S,n)$ for a single source.
The sources are assumed to be independent. This gives at $i\neq j$:
\eq{\label{ind-source}
\overline{X_iX_j}~\equiv ~\overline{X_S}^2~,~~~~
\overline{n_in_j}~\equiv ~\overline{n}^2~,~~~
\overline{X_in_j}~\equiv ~\overline{X_S}~\overline{n}~.
}
Using Eqs.~(\ref{id-source}) and (\ref{ind-source})
one finds for the event averages:
 \eq{
 %
 &\langle X \rangle~=~\overline{X_S}\cdot\langle N_S\rangle~,~~~~
\langle X^2 \rangle~=~\overline{X_S^2}\cdot \langle N_S \rangle ~+~
 \overline{X_S}^2
 ~\left[\langle N_S^2\rangle ~-~\langle N_S\rangle\right]~,\label{WNM-X2}\\
&\langle N \rangle~=~\overline{n}~\langle N_S\rangle~,~~~~
\langle N^2 \rangle~=~\overline{n^2}\cdot \langle N_S \rangle ~+~
 \overline{n}^2\cdot
\left[\langle N_S^2\rangle ~-~\langle N_S\rangle\right]~,\label{WNM-N2} \\
 &\langle X\,N \rangle~=~
\overline{X_S\,n} ~\langle N_S \rangle ~+~
 \overline{X_S}~\overline{n}\cdot
 \left[\langle N_S^2\rangle ~-~\langle
 N_S\rangle\right]~.\label{WNM-XN}
}
A probability distribution ${\cal P}_S(N_S)$ of the number of
sources is needed to calculate $\langle N_S\rangle $ and $\langle
N_S^2\rangle$ and, in general, it is unknown.

Using Eqs.~(\ref{WNM-X2}-\ref{WNM-XN}) one obtains:
 \eq{\label{WNM-omega-X}
  \omega[X]&\equiv ~\frac{\langle X^2\rangle
-\langle X \rangle ^2}{\langle X\rangle}=\frac{
\overline{X_S^2}-\overline{X_S}^2}{\overline{X_S}}+\overline{X_S}
\cdot \frac{\langle N_S^2\rangle -\langle
N_S\rangle^2}{\langle N_S\rangle} ~\equiv~\omega[X_S]~+~\overline{X_S}\cdot
\omega[N_S]~,\\
%
%
\label{WNM-omega-N}
  \omega[N]&\equiv~ \frac{\langle N^2\rangle
~-~\langle N \rangle ^2}{\langle N\rangle}~=~\frac{
\overline{n^2}~-~\overline{n}^2}{\overline{n}}~+~\overline{n}
\cdot \frac{\langle N_S^2\rangle ~-~\langle
N_S\rangle^2}{\langle N_S\rangle} ~\equiv~\omega[n]~+~\overline{n}\cdot
\omega[N_S]~,
}
where $\omega[X_S]$ and $\omega[n]$ are the scaled variances for
quantities $X_S$ and $n$ referring to a single source.  The scaled
variances $\omega[X]$ and $\omega[N]$ are independent of the
average number of sources $\langle N_S\rangle$. Thus, $\omega[X]$
and $\omega[N]$ are intensive quantities. However, they depend on
the fluctuations of the number of sources  via $\omega[N_S]$ and,
therefore, they are not strongly intensive quantities.

From Eqs.~(\ref{WNM-XN}-\ref{WNM-omega-N}) it follows:
 \eq{
\Delta[X,N]~&=~\frac{1}{\omega[x]}~
\Big[\, \omega[X_S] ~-
~
\overline{x}\cdot \omega[n] \,\Big]~. \label{D-WNM}\\
\Sigma[X,N]~&=~
\frac{1}{\omega[x]}~
\Big[\,
\omega[X_S] ~+~
\overline{x}\cdot \omega[n]~-~
2\,\frac{\overline{X_S\, n} -\overline{x}~\overline{n}^2}{\overline{n}}\,\Big]~,
\label{S-WNM}
}
where the relations
%
%
$ \overline{x}= \overline{X_S}/\overline{n}=\langle X\rangle/\langle N\rangle$~,
%
%
and the normalization factors (\ref{norm})
have been used.

Note that the terms with $\langle N_S^2\rangle$, which are present
in the expressions (\ref{WNM-X2}-\ref{WNM-XN}) for the second
moments of $X$ and $N$, are canceled out in the final expressions
(\ref{D-WNM},\ref{S-WNM}). From three second moments $\langle
X^2\rangle$, $\langle N^2\rangle$, and $\langle X\,N\rangle$ only
two linear combinations independent of $\langle N_S^2\rangle$ can
be constructed. They are defined as the strongly intensive
quantities $\Delta$ and $\Sigma$. To remove the dependence on
$\langle N_S\rangle$, the strongly intensive quantities should be
in a form of reducible fractions. This is achieved due to the
normalization factors (\ref{norm}).

Only the first and second moments of $X$ and $N$ are required in
order to define the strongly intensive quantities $\Delta$ and
$\Sigma$. However, in order to calculate the proposed
normalization factors $C_{\Sigma}$ and $C_{\Delta}$, additional
information is needed, namely the second moment $\overline{x^2}$
of single-particle distribution $F(x)$. Note that the first moment
$\overline{x}$ can be calculated as $\overline{x}=\langle X\rangle
/\langle N\rangle$, and thus to find it the single particle
distribution $F(x)$ is not necessarily needed.

The IPM and MIS  have  similar structure. The difference is that
the number of sources $N_S$ in  the MIS is replaced by the number
of particles $N$ in  the IPM. Each source can produce many
particles, and the number of these particles varies from source to
source and from event to event. Besides, the physical quantity
$X_S$ for particles emitted from the same source may include
inter-particle correlations. Therefore, in general,  the MIS does
not satisfy the assumptions of  the IPM. Nevertheless, a formal
similarity between the two models can be exploited and gives the
following rule of one to one correspondence: all results for
the IPM can be found from the expressions obtained within the MIS,
assuming artificially that each source always produces exactly one
particle. In this case one finds
\eq{
\overline{n}=1~,~~~~\omega[n]=0~,~~~~\omega[X_S]=\omega[x]~,~~~~
\overline{X_S\,n}=\overline{x}~,
}
and Eqs.~(\ref{D-WNM}-\ref{S-WNM}) are transformed to Eq.~(\ref{DelSig=1}).

If particles are independently emitted from a single source, one
obtains
\eq{\label{PSx}
F_S(X_S,n)~=~{\cal P}_S(n)\times F_S(x_1)\times \dots \times F_S(x_n)~,
}
with the probability distributions  $F_S(x_i)$ which are the same
for all $i=1,\ldots, n$ and independent of the number of particles
$n$. Similar to Eqs.~(\ref{omega-X-IPM}) and (\ref{XN-IPM}) one
then finds:
\eq{
 \omega[X_S]~=~\omega[x]~+~\overline{x}\cdot \omega[n]~,~~~~
 \overline{X_S\,n}~-~\overline{X_S}\,\overline{n}~=~\overline{x}~\overline{n}\cdot \omega[n]~,
}
and Eqs.~(\ref{D-WNM}) and (\ref{S-WNM}) are again transformed to
Eq.~(\ref{DelSig=1}). Therefore,  the MIS with independent particle
emission from each source is equivalent to the IPM.

Correlations of particles emitted from a single source can be of
different origin. Let us consider the case when all single-particle
distributions $F_S(x)$ in Eq.~(\ref{PSx}) are dependent on the
source parameter $T$ (e.g., the source temperature) which
fluctuates, and these $T$-fluctuations are independent for each
source. The $F_S(X_S,n)$ distribution for a single source can be
then presented as
\eq{\label{no-corr}
F_S(X_S,n)~=~{\cal P}_S(n)\times \int dT\,W(T)~ F_S(x_1,T)\times \dots \times F_S(x_n,T)~.
}
Note that presentation (\ref{no-corr}) means the absence of
correlations between particle momenta $x_j$ and multiplicity $n$,
but correlations between $x_i$ and $x_j$ appear due to the $T$
fluctuations. The multi-particle distribution (\ref{no-corr}) may
look as a simple product of the one-particle distributions.
However, the single particle distributions are not independent
due to integration over $T$. With distribution
(\ref{no-corr}) one calculates
\eq{
\overline{X_S\,n}&=\sum_n {\cal P}_S(n)
\int dT W(T)\int dx_1\dots dx_n F_S(x_1,T) \dots F_S(x_n,T)\,
(x_1+\dots+x_n)\cdot n
\nonumber ~\\
 &=\overline{x}\,\overline{n^2}~.\label{XSn}
}
Using Eq.~(\ref{XSn}) one can simplify further Eq.~(\ref{S-WNM}).
Finally, it gives:
\eq{
\label{DSWNM}
~\Delta[X,N]~=\Sigma[X,N]~=~\frac{\omega[X_S]~-~\overline{x}\cdot \omega[n]}
{\omega[x]}~.
}

\section{Fast Generators and Analytical Results}\label{fast}
We consider the Boltzmann {\it transverse} momentum ($p\equiv p_T$) distribution
\eq{\label{boltz}
f(p,T)~=~C\,p\,\exp\Big(-~\frac{\sqrt{m^2+p^2}}{T}\Big)~,
}
where constant $C$ is defined by the normalization condition and
%
%
$C^{-1}=\int_{0}^{\infty}dpp\,\exp(-\sqrt{m^2+p^2}/T)$~.
%
%
The particle mass $m$ in the MC simulations is taken as the pion mass
$m=m_\pi\cong 140$~MeV, $T$ is the {\it effective} temperature or
simply an inverse slope parameter controlled by the actual
freeze-out temperature and the collective transverse flow
velocity.
The moments ($k=1,2$) of the $f(p,T)$  probability distributions
(\ref{boltz}) are denoted as
 \eq{
 \tilde{p^k}~&=~\int_0^{\infty}dp~p^k~f(p,T)~.
%
%
 \label{p2}
}
In the presence of e-by-e temperature fluctuations, the inclusive
transverse momentum distributions reads
\eq{\label{fp}
f(p)~=~\int dT\,W(T)~f(p,T)~,
}
where
$W(T)$ is the temperature probability distribution normalized to
one. The moments ($k=1,2$) of the $f(p)$ probability distribution
(\ref{fp}) are denoted as
 \eq{
 \overline{p^k}~=~\int_0^{\infty}dp~p^k~f(p)~=~\int dT \,W(T)~\tilde{p^k}~.
 \label{pp2}
}

In the case of massless particles $m=0$  distribution (\ref{boltz})
is reduced to a simple exponential form and one can easily compute
\eq{\label{pm0}
 \tilde{p}=2T~,~~~~\tilde{p^2}=6T^2~,~~~~\overline{p}=2\overline{T}~,
 ~~~~\overline{p^2}=6\overline{T^2}~,~~~~
 \omega[p]=\frac{3\overline{T^2}-2\overline{T}^2}{\overline{T}}~,
}
where ($k=1,2$)
\eq{\label{Tk}
\overline{T^k}~=~\int dT\,T^k~W(T)~.
}
Note that in the MC simulations the particle transverse  momenta
are generated with the $p$-values in a region $[0,p_{{\rm max}}]$.

The basic properties of $\Delta[P_T, N]$ and $\Sigma[P_T, N]$
measures will be tested using  MC simulations
(so-called fast generators). Each interaction (event) is composed
by a given number of sources. For each simulation the
statistical errors on $\Delta[P_T, N]$ and $\Sigma[P_T, N]$ are
estimated as follows. The whole sample of events is divided into
30 independent sub-samples. Next, the values of $\Delta[P_T, N]$
and $\Sigma[P_T, N]$ are evaluated for each sub-sample and the
dispersions ($D_{\Delta}$, and $D_{\Sigma}$) of the results are
then calculated. The statistical error of $\Delta[P_T, N]$ or
$\Sigma[P_T, N]$ is taken to be equal to $D_{\Delta}/\sqrt{30}$ or
$D_{\Sigma}/\sqrt{30}$, respectively.

\subsection{Fixed Temperature}
\begin{figure}
\centering
\includegraphics[width=0.45\textwidth]{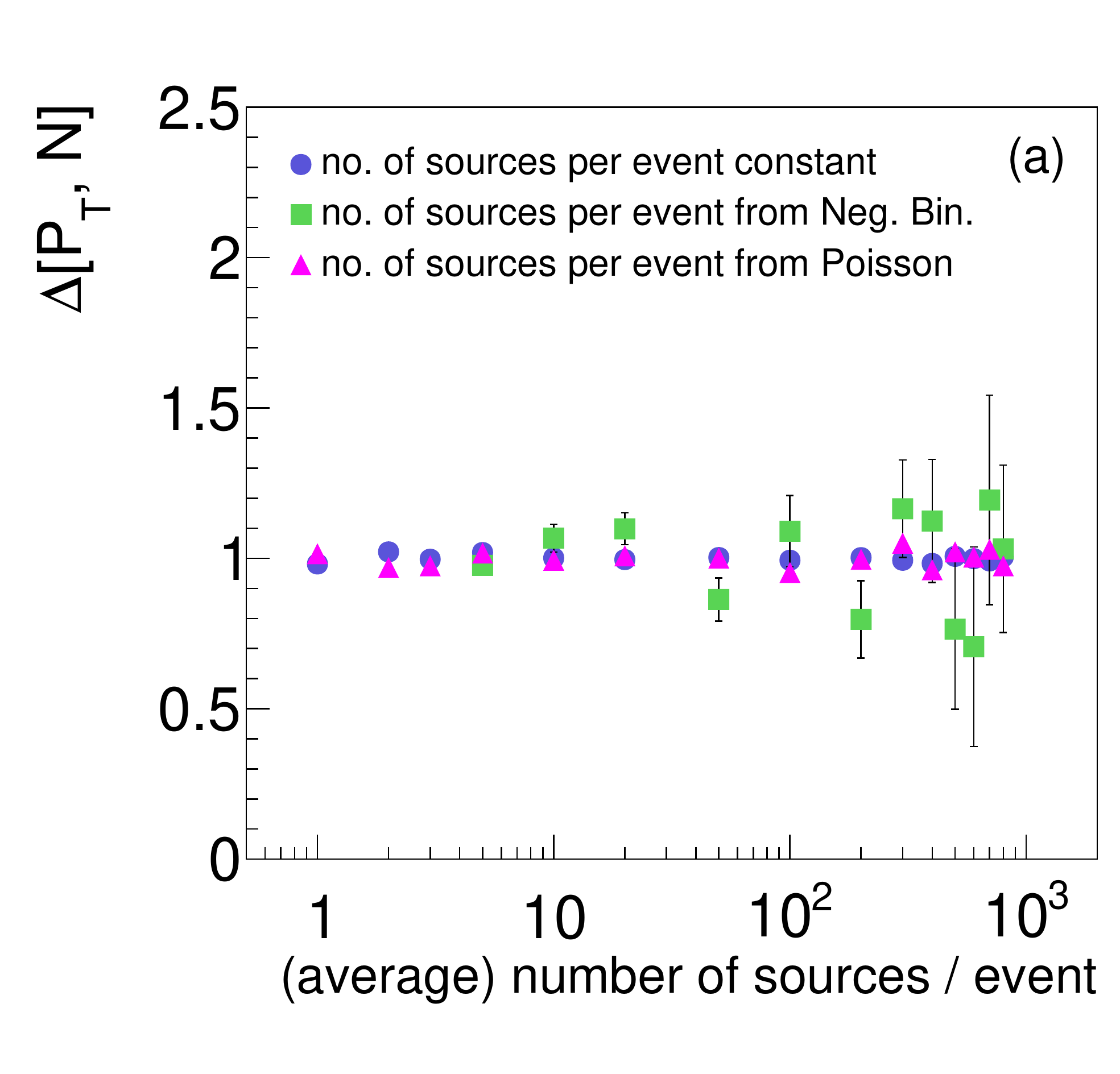}
\includegraphics[width=0.45\textwidth]{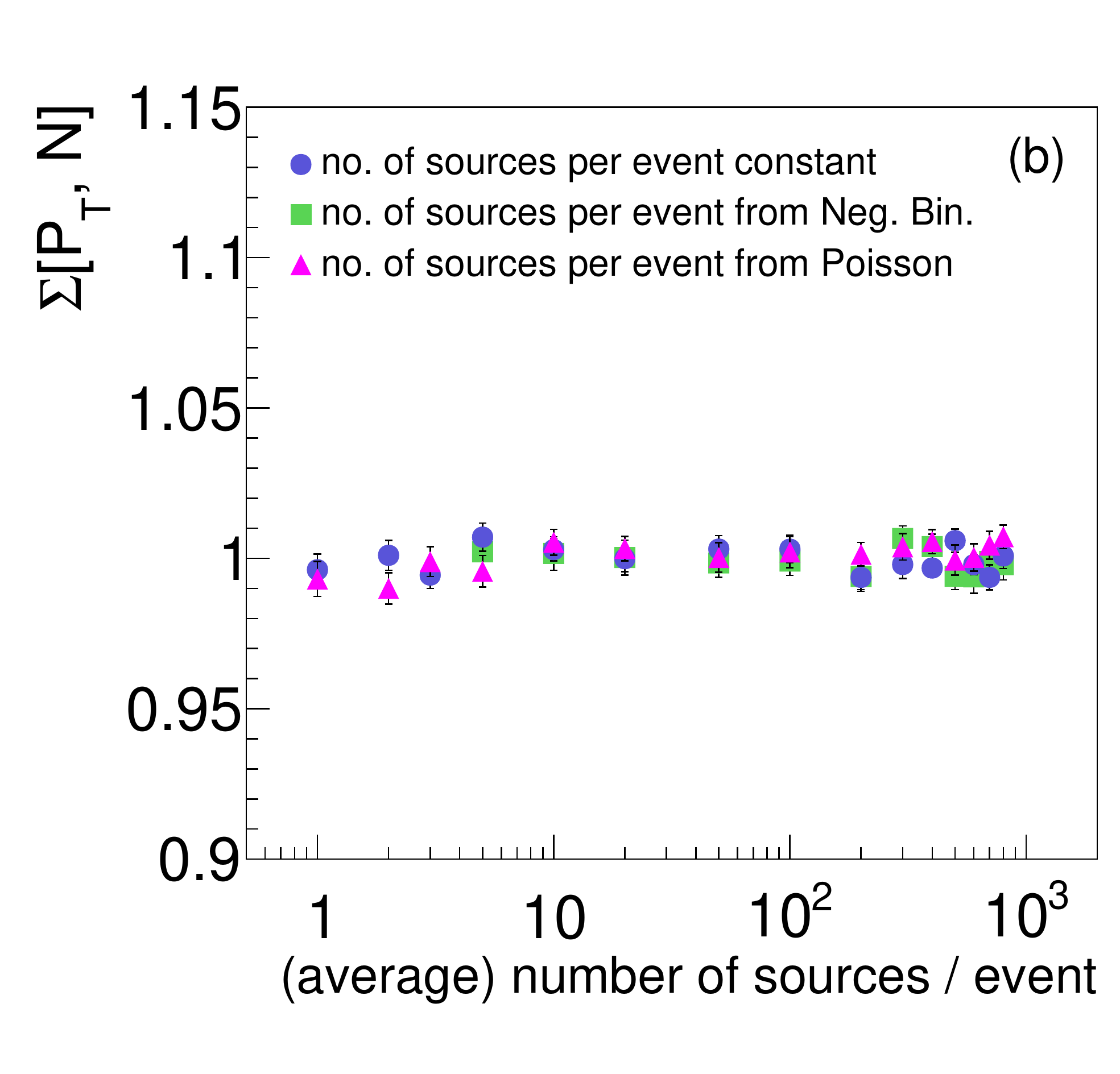}
\caption[]{(Color online) The symbols are the MC results for
the $\Delta[P_T, N]$ ({\it a}) and
$\Sigma[P_T, N]$ ({\it b}) measures versus the number or
mean number of sources composing one event. All sources
have fixed temperature. The number of sources per event are
fixed ({\it circles}) or fluctuating according to
the Poisson distribution ({\it triangles}) and Negative Binomial distribution
({\it squares}).}
\label{figfixT}
\end{figure}

The first set of the MC simulations refers to particle production
from sources with  fixed temperature. For each source in a given
event the number of particles was generated from the Poisson
distribution with a mean value of 5. The particle transverse
momentum was generated from transverse momentum distribution
(\ref{boltz}) with maximal value $p_{{\rm max}}=2.0$~GeV/$c$.
The temperature parameter is fixed as $T=150$~MeV. The number of
sources $N_S$ composing an event is either constant (circles in
Fig.~\ref{figfixT}) or selected from Poisson (triangles) or from
Negative Binomial distribution (squares). For Negative Binomial
distribution its dispersion $\sqrt{Var(N_{S})}$ is large and
equals $\langle N_{S} \rangle / 2$.

Figure \ref{figfixT} shows  $\Delta[P_T, N]$ and $\Sigma[P_T,
N]$  versus the number or the mean number of sources
composing one event.
The distribution function of a single source has the form of
Eq.~(\ref{PSx}) and the $P_S(x)$ function is taken as $f(p,T)$
(\ref{boltz}) with fixed temperature $T$, same for all
sources. This corresponds the case when the MIS is reduced to
the IPM, and relation (\ref{DelSig=1}) should be valid. As expected,
the $\Delta[P_T, N]$ and $\Sigma[P_T, N]$ values for the MC
simulations are consistent with {\it one}, independently of the
assumed shape of transverse momentum distribution.
The circles in Fig.~\ref{figfixT} confirm that $\Delta[P_T, N]$
and $\Sigma[P_T, N]$ are intensive measures (do not depend on
$N_S$), whereas the triangles and the squares show that these
quantities are also strongly intensive (do not depend on $N_S$
fluctuations).

For a constant number of sources per event (circles in
Fig.~\ref{figfixT}), the scaled variance of multiplicity
distribution $\omega[N]=\omega[n]+\overline{n}\omega[N_S]$  equals
1 in the whole range of the horizontal axis. For the Poisson
distribution of the number of sources $\omega[N]$ equals to 6
also for the whole range of the mean number of sources per event. For
the Negative Binomial distribution of the number of sources
$\omega[N]$ increases from about 7 at $\langle N_{S} \rangle=5$,
through 126 at $\langle N_{S} \rangle=100$, up to approximately 1000
at $\langle N_{S} \rangle=800$. Therefore, Fig.~\ref{figfixT} shows
that $\Delta[P_T, N]$ and $\Sigma[P_T, N]$ measures are strongly
intensive even for multiplicity distributions which are extremely
wide.

\subsection{Source Temperature Fluctuations}
\begin{figure}
\centering
\includegraphics[width=0.65\textwidth]{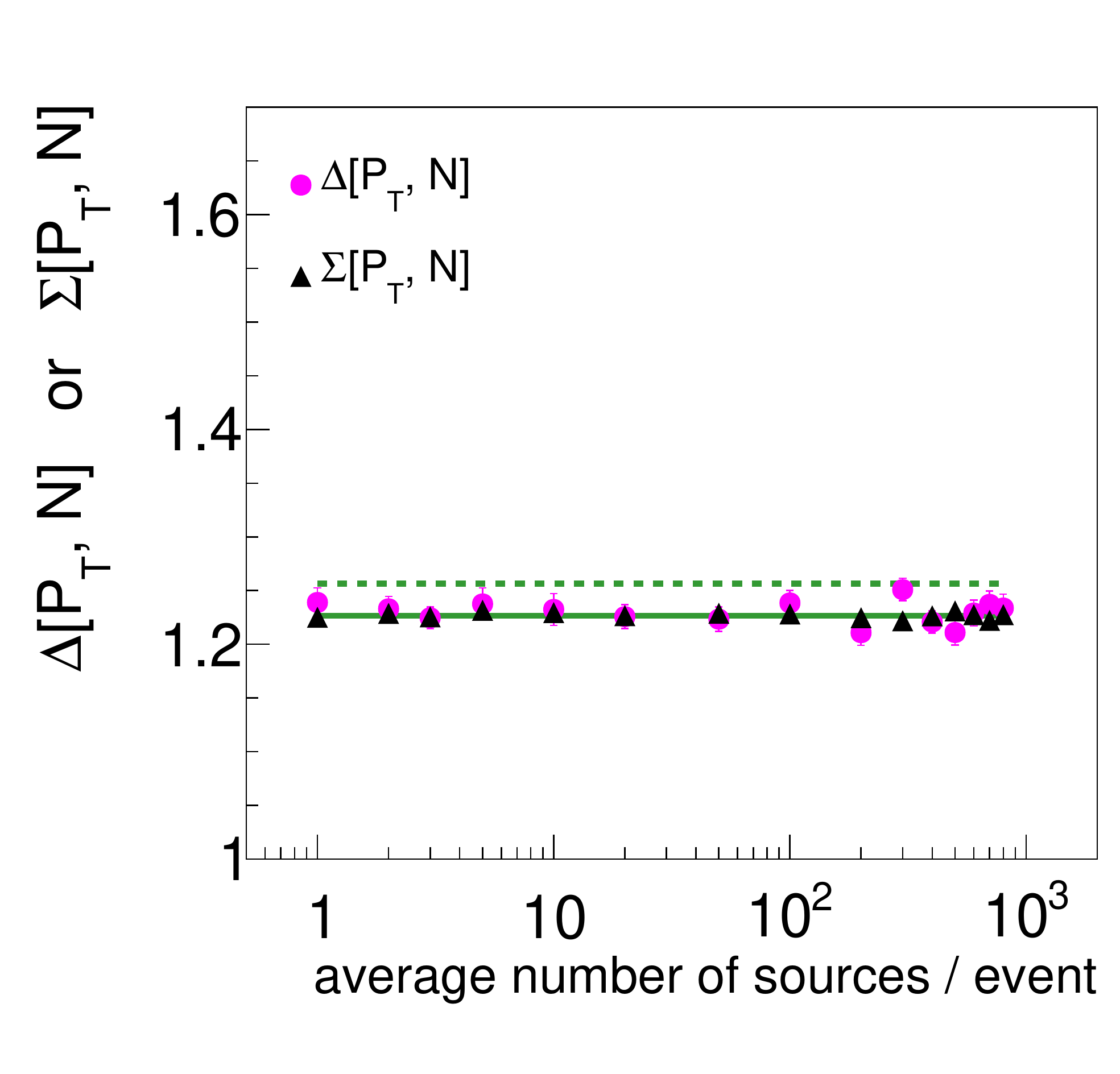}
\caption[]{(Color online) The symbols are the MC results for the $\Delta[P_T, N]$
({\it circles}) and $\Sigma[P_T, N]$ ({\it triangles}) measures
versus the mean number of sources composing one event. The
temperatures of the sources fluctuate independently according to
Eq.~(\ref{WT}), and the number of sources per event fluctuates
according to the Poisson distribution. The solid line corresponds
to Eq.~(\ref{DS-Tfluc-boltz}), the dashed line to
Eq.~(\ref{DS-m0}).} \label{figsbys}
\end{figure}

In the next set of simulations,  the number of particles produced
by each single source is again selected from the Poisson
distribution with a mean value of $\overline{n}=5$. The particle
transverse momentum is generated by the transverse momentum
distribution (\ref{boltz}) with average inverse slope parameter
$\overline{T} =150$~MeV. The $T$ parameter is generated separately
for each single source (source-by-source $T$ fluctuations) from
the Gaussian distribution
\eq{\label{WT}
W(T)~ = ~\frac{1}{\sqrt{2\pi}\,\sigma_T} \exp \Big[\, -~ \frac
{(T-\overline{T})^2} {2\sigma_T^2}  \Big]~,
}
with dispersion $\sigma_{T}=25$ MeV. Finally, the number of
sources $N_S$ composing an event  is generated from the Poisson
distribution, with $\langle N_{S} \rangle$ as denoted on the
horizontal axis of Fig.~\ref{figsbys}. As seen, the effect of
source temperature fluctuations results in $\Delta[P_T, N]$ and
$\Sigma[P_T, N]$ values higher than 1.

If the parameter $T$ fluctuates {\it independently} for each
source, the sources remain to be statistically identical and
independent of each other.  Therefore, these MC simulations
correspond to the MIS and  the $\Delta[P_T, N]$ and $\Sigma[P_T,
N]$ strongly intensive measures should not depend on the mean
number of sources $\langle N_S\rangle$ and on its fluctuations
$\omega[N_S]$. Indeed, Fig.~\ref{figsbys} confirms this
expectation.

The distribution function of a single source has the form of
Eq.~(\ref{no-corr}) with $F_S(x,T)$ function taken as $f(p,T)$.
This leads to the result ($P\equiv P_T$) (\ref{DSWNM})
\eq{\label{DSTfluc}
\Delta[P,N]~=~\Sigma[P,N]~=~\frac{\omega[P_S]~-~\overline{p}\cdot
\omega[n]} {\omega[p]}~,~~~~\omega[p]~\equiv~\frac{ \overline{p^2}
~-~\overline{p}^2}{ \overline{p}}.
}
For $\overline{P_S}$ and $\overline{P_S^2}$ one obtains:
 \eq{
 & \overline{P_S}~=~\overline{p}\cdot \overline{n}
 ~, \label{PS}\\
 & \overline{(P_S)^2}~=~\sum_n {\cal P}_S(n)~\int dT W(T)~
 \int_0^{\infty}\prod_{i=1}^n\Big[p_idp_i\, f(p_i)\Big]~
 (p_1+\ldots +p_n)^2~\nonumber\\
 &=~\sum_n {\cal P}_S(n)~\int dT W(T)~
 \int_0^{\infty}\prod_{i=1}^n\Big[p_idp_i\,f(p_i)\Big]~
 \Big[\sum_{j=1}^n p_j^2~+\sum_{1\leq l\neq m\leq k} p_l\cdot p_m \Big] \nonumber\\
 &=~ \sum_n {\cal P}_S(k)~\int dT W(T)~\Big[n\cdot \tilde{p^2}~+~n(n-1)\cdot \tilde{p}^2\Big]
 ~ =~\overline{p^2}\cdot \overline{n}~+~\hat{p}^2\cdot\Big[ \overline{n^2}~-~\overline{n}\Big]~,\label{PS2}
}
where
\eq{\label{phat}
\hat{p}^2~\equiv~\int dT\,W(T)~\tilde{p}^2~.
 }

Calculating $\omega[P_S]$ from Eqs.~(\ref{PS},\ref{PS2}) and
inserting it into Eq.~(\ref{DSWNM}) one obtains:
\eq{\label{DS-Tfluc}
\Delta[P,N]~=~\Sigma[P,N]~
=~ 1~+~\frac{1}{\omega[p]}\cdot
\frac{\hat{p}^2~-~\overline{p}^2}{\overline{p}}\cdot
\Big[\overline{n}~+~\omega[n]~-~1\Big]~.
}
%
%
%
One can easily prove that
\eq{ & \hat{p}^2~-~\overline{p}^2~=~\int dT\,W(T)~(\tilde{p}~-
~\overline{p})^2~\geq~0~, \label{pp}\\
& \overline{n}~+~\omega[n]~-~1~=~\frac{\overline{n^2}~
-~\overline{n}}{\overline{n}}~=~\frac{1}{\overline{n}}~\sum_{n\geq
2} {\cal P}_S(n)\,(n^2~-~n)~ \geq ~0~.\label{ng2}
}
When temperature fluctuations are absent, relation~(\ref{pp})
is transformed to $\hat{p}^2-\overline{p}^2=0$, and
Eq.~(\ref{DS-Tfluc}) is reduced to Eq.~(\ref{DelSig=1}). The same
happens when ${\cal P}_S(n)=0$ for all $n\geq 2$, and, thus,
$\overline{n}+\omega[n]-1=0$.  This is intuitively clear: the MIS
is reduced to  the IPM if each source can emit only one or zero
number of particles.

In Fig.~\ref{figsbys} the results of the MC  simulations are compared
with analytical results of Eq.~(\ref{DS-Tfluc}). The solid line
corresponds to the distribution
(\ref{boltz}) and Gaussian temperature fluctuations (\ref{WT}).
In this case one finds:
\eq{\label{pp-m}
 \overline{p}\cong 0.328~{\rm GeV}/c~,~~~~  \overline{p^2}\cong 0.158~({\rm GeV}/c)^2~,~~~~
\hat{p}^2\cong 0.110~({\rm GeV}/c)^2~.
}
The ${\cal P}_S(n)$ Poisson distribution for a single source
corresponds to $\overline{n}=5$ and $\omega[n]=1$, therefore,
$\overline{n}+\omega[n]-1=5$.
The final result of Eq.~(\ref{DS-Tfluc}) is
\eq{\label{DS-Tfluc-boltz}
\Delta[P,N]~=~\Sigma[P,N]~\cong 1.227~.
%
}
As seen in Fig.~\ref{figsbys},
this is in a good agreement with the results of the MC simulations.

For massless particles the quantities in Eq.~(\ref{pp-m})
can be calculated analytically
\eq{\label{pp-m0}
&
\overline{p}~=~2\overline{T}~=~0.3~{\rm GeV}/c~,~~~~
 ~~~~\overline{p^2}=6\overline{T^2}~=~6\,[\overline{T}^2~+~\sigma^2_T]~=~0.13875~({\rm GeV}/c)^2 ,~~~~
\\
& \hat{p^2}~=~\int
dT\,W(T)\,\tilde{p}^2~=~4\overline{T^2}~=4[\overline{T}^2~+~\sigma_T^2]~
=~0.0925~({\rm GeV}/c)^2~.
\label{p2-hat}
}
With Eqs.~(\ref{pp-m0},\ref{p2-hat}) one finds
\eq{\label{DS-m0}
\Delta[P,N]~=~\Sigma[P,N]~
=~ 1~+~\frac{2\sigma_T^2}{\overline{T}^2+3\sigma_T^2}\cdot
\Big[\overline{n}~+~\omega[n]~-~1\Big]~.
}
For the values $\overline{T}=0.15$~GeV, $\sigma_T=0.025$~GeV,
$\overline{n}=5$, and $\omega[n]=1$ used in the MC simulations
one finds $\Delta[P,N]=\Sigma[P,N]\cong 1.256$. This result for
$m=0$ is shown in Fig.~\ref{figsbys} by the dashed line.

The MC results presented in
Figs.~\ref{figfixT} and \ref{figsbys} demonstrate a different
sensitivity of the strongly intensive measures to model details:
despite of the equality $\Delta[P_T,N]=\Sigma[P_T,N]$ the statistical
errors of the simulations calculated for $\Delta[P_T,N]$ are
found to be essentially larger than those for $\Sigma[P_T,N]$ \footnote{In order to avoid too large statistical errors, in Fig.~\ref{figsbys} we used five times higher statistics (500k events for each point) than that one used in Figs.~\ref{figfixT} and \ref{figebye}.}.

\subsection{Global Temperature Fluctuations}
\begin{figure}
\centering
\includegraphics[width=0.49\textwidth]{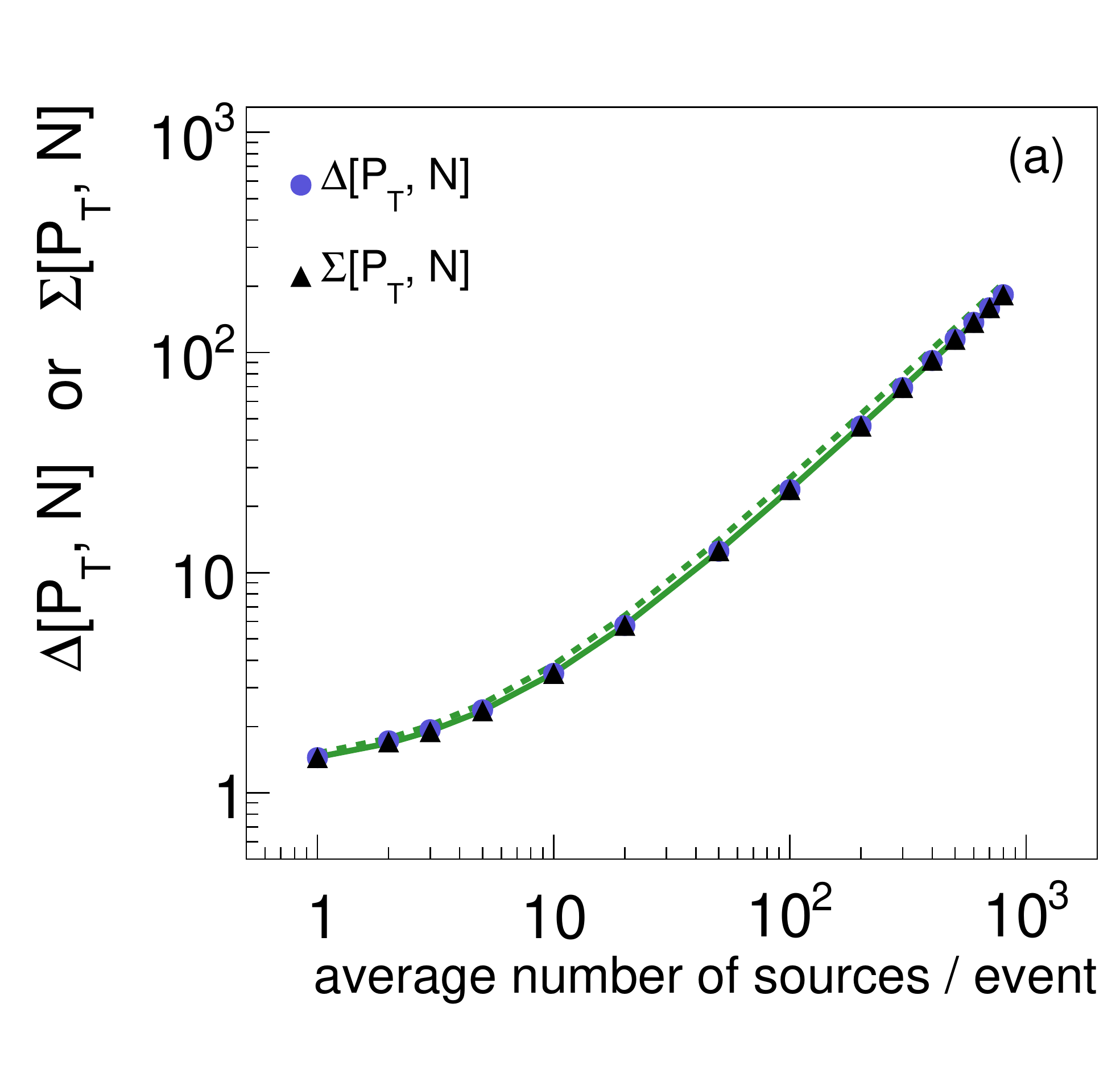}
\includegraphics[width=0.49\textwidth]{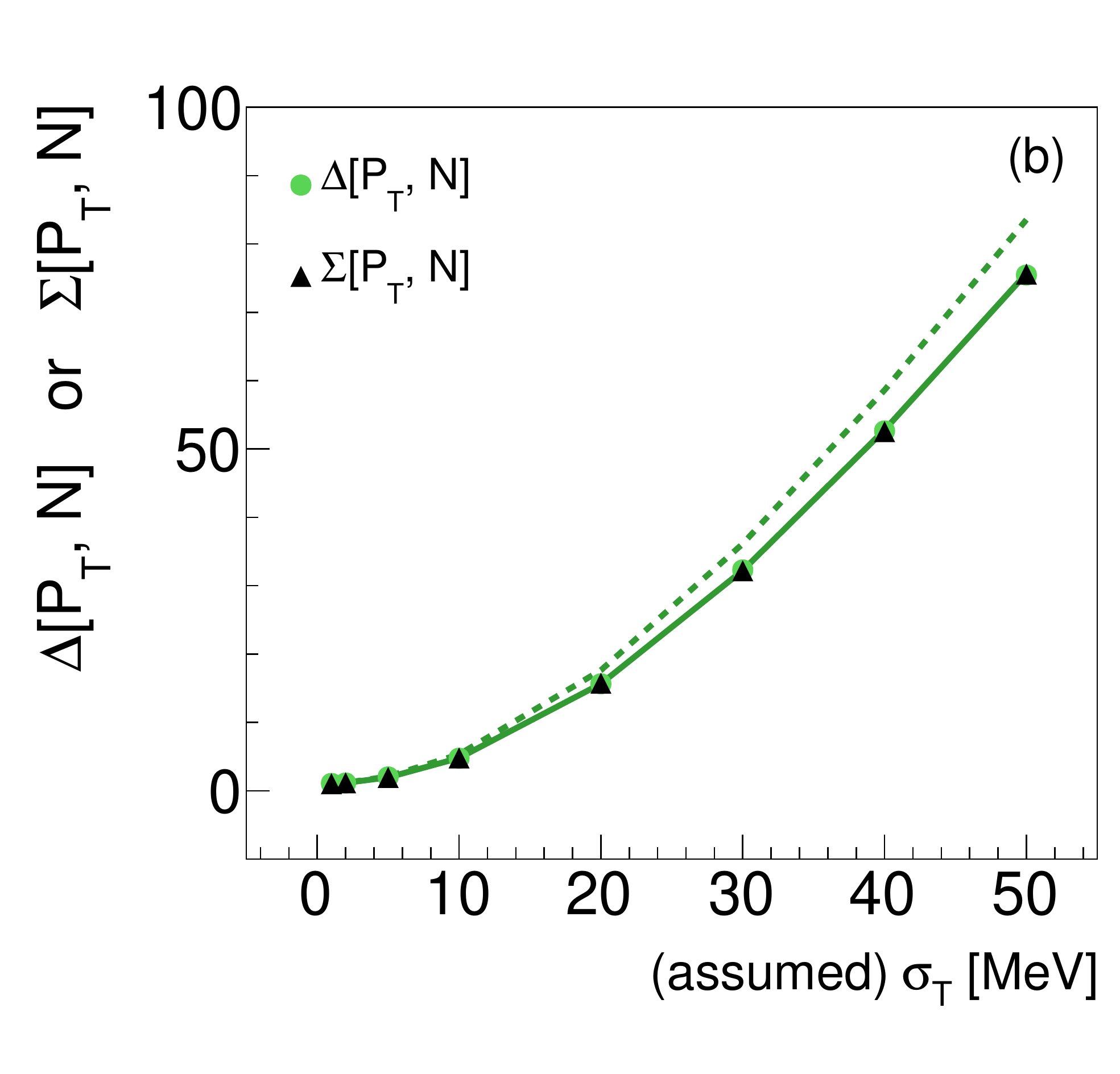}
\caption[]{(Color online) The symbols are the MC results for the
 $\Delta[P_T, N]$ ({\it circles}) and $\Sigma[P_T, N]$ ({\it triangles})
measures.  The MC
simulations correspond to the {\it global} temperature fluctuations according to
Eq.~(\ref{WT}), i.e.  temperatures of all sources are equal. The number of sources
are taken from the Poisson distribution with the average value of $\langle N_S\rangle$.
The solid lines present the results of Eq.~(\ref{DS-Tfluc-global}),
the dashed lines of Eq.~(\ref{DS-m00}).
({\it a}): The dependence on the mean number of sources at fixed $\sigma_T=25$~MeV.
({\it b}): The dependence on $\sigma_T$ at fixed $\langle N_S\rangle =100$ (here for calculating dashed line the {\it obtained} $\sigma_T$ values were used; due to the limited range of $T$ distribution they are slightly smaller than the {\it assumed} ones).}
\label{figebye}
\end{figure}

In the next MC simulations, source-by-source $T$ fluctuations from
the previous subsection are replaced by e-by-e (global) $T$
fluctuations. The parameter $T$ is the same for all sources
composing a given event but is varied between events following the
Gaussian distribution (\ref{WT}) with average
inverse slope parameter $\langle T\rangle=150$~MeV and dispersion
$\sigma_{T}$.
The number of sources $N_S$ composing an event is generated
from the Poisson distribution with $\langle N_{S} \rangle$ being the
average value. As previously, for
each single source, the number of particles was selected from the
Poisson distribution with a mean value of $\overline{n}=5$.
The results are presented in Fig.~\ref{figebye}.
Its  ({\it a}) panel shows the dependence of
$\Delta[P_T,N]$ and $\Sigma[P_T, N]$) on
the average number of sources $\langle N_S\rangle$ at $\sigma_T=25$~MeV,
whereas the ({\it b}) panel
presents the dependence on $\sigma_T$ at $\langle N_S\rangle=100$.
In Fig.~\ref{figebye} ({\it b}) in order to avoid negative $T$ values only events
within $T=150 \pm 3 \sigma_{T}$~MeV were accepted.
We also would like to mention here that the
relationship between temperature and multiplicity (or volume)
fluctuations was studied in Refs.~\cite{volume,multip_T_fluct}.

Due to the
correlated $T$-fluctuations for different sources, the sources
{\it are not} independent of each other. Therefore, these MC
simulations {\it do not} correspond to the MIS.
%
One can nevertheless use the formula from the previous subsection
with the following substitutions:
\eq{
N_S\rightarrow 1~,~~~~\omega[N_S]\rightarrow 0~,~~~~ n\rightarrow
N~,~~~~
P_S\rightarrow P ~, 
}
i.e. all final particles are treated as created from a ``single source''
with fluctuating temperature $T$. Note that the parameter $T$ becomes
an event variable with average value $\langle T\rangle=\overline{T}=150$~MeV and
distribution (\ref{WT}).
This gives:
\eq{\label{DS-Tfluc-global}
\Delta[P,N]~=~\Sigma[P,N]~=~ 1~+~\frac{1}{\omega[p]}\cdot
\frac{\hat{p}^2~-~\overline{p}^2}{\overline{p}}\cdot
\Big[\,\langle N\rangle ~+~\omega[N]~-~1\,\Big]~.
}

The MC
results on global temperature fluctuations are compared to
analytical predictions of Eq.~(\ref{DS-Tfluc-global}). The solid
lines in Fig.~\ref{figebye} correspond to the transverse momentum
distribution (\ref{boltz}) with temperature fluctuating according
to Eq.~(\ref{WT}).
The values of $\overline{p}$, $\overline{p^2}$, $\hat{p}^2$,
and $\omega[p]$
are
calculated numerically with Eqs.~(\ref{boltz}) and (\ref{WT}).
At $\sigma_T=25$~MeV, they are equal to those in Eq.~(\ref{pp-m}).
Analytical calculations can be done for massless particles
according to Eq.~(\ref{pp-m0}) which demonstrates the explicit
dependence on $\sigma_T$.

Note that
multiplicities $n_1,\dots, n_{N_S}$ for particles emitted by different
sources are uncorrelated. Therefore, one can use the MIS to
calculate $\langle N\rangle$ and  $\omega[N]$ with
Eq.~(\ref{WNM-omega-N}):
\eq{\label{omegaN}
\langle N \rangle = \overline{n} \langle N_S \rangle = 5 \langle
N_S \rangle~,~~~~
\omega[N]=\omega[n]+\overline{n} \cdot \omega[N_S]=1+5 \cdot 1=6~.
}
This results in
a linear increase of (\ref{DS-Tfluc-global}) with $\langle N_S\rangle$.

For $m=0$ in the distribution (\ref{boltz}),   similarly to
Eq.~(\ref{DS-m0}), one obtains
\eq{\label{DS-m00}
\Delta[P,N]~=~\Sigma[P,N]~
=~1~+~\frac{2\sigma_T^2}{\langle T\rangle^2+3\sigma_T^2}\cdot
\Big[5\, \langle N_s\rangle ~+~5\Big]~,
}
where Eq.~(\ref{omegaN}) has been already used.
This  is shown in Fig.~\ref{figebye} by dashed lines.

%

As expected from
Eq.~(\ref{DS-Tfluc-global}), the fluctuation measures $\Delta[P_T,
N]$ and $\Sigma[P_T, N]$) increase when global temperature
fluctuations are stronger (higher $\sigma_{T}$). This is
explicitly seen from Eq.~(\ref{DS-m00}) for $m=0$.  The same
conclusion was drawn in Ref.~\cite{m8}, where the influence of
temperature fluctuations on transverse momentum fluctuations was
studied for the $\Phi_{p_{T}}$  measure \cite{GM:1992} (see also
Ref.~\cite{KG:APP2012} for the corresponding plot).

%

\section{Temperature Correlations Versus Number of Particles}\label{TN-corr}

\begin{figure}[ht]
\centering
\includegraphics[width=0.95\textwidth]{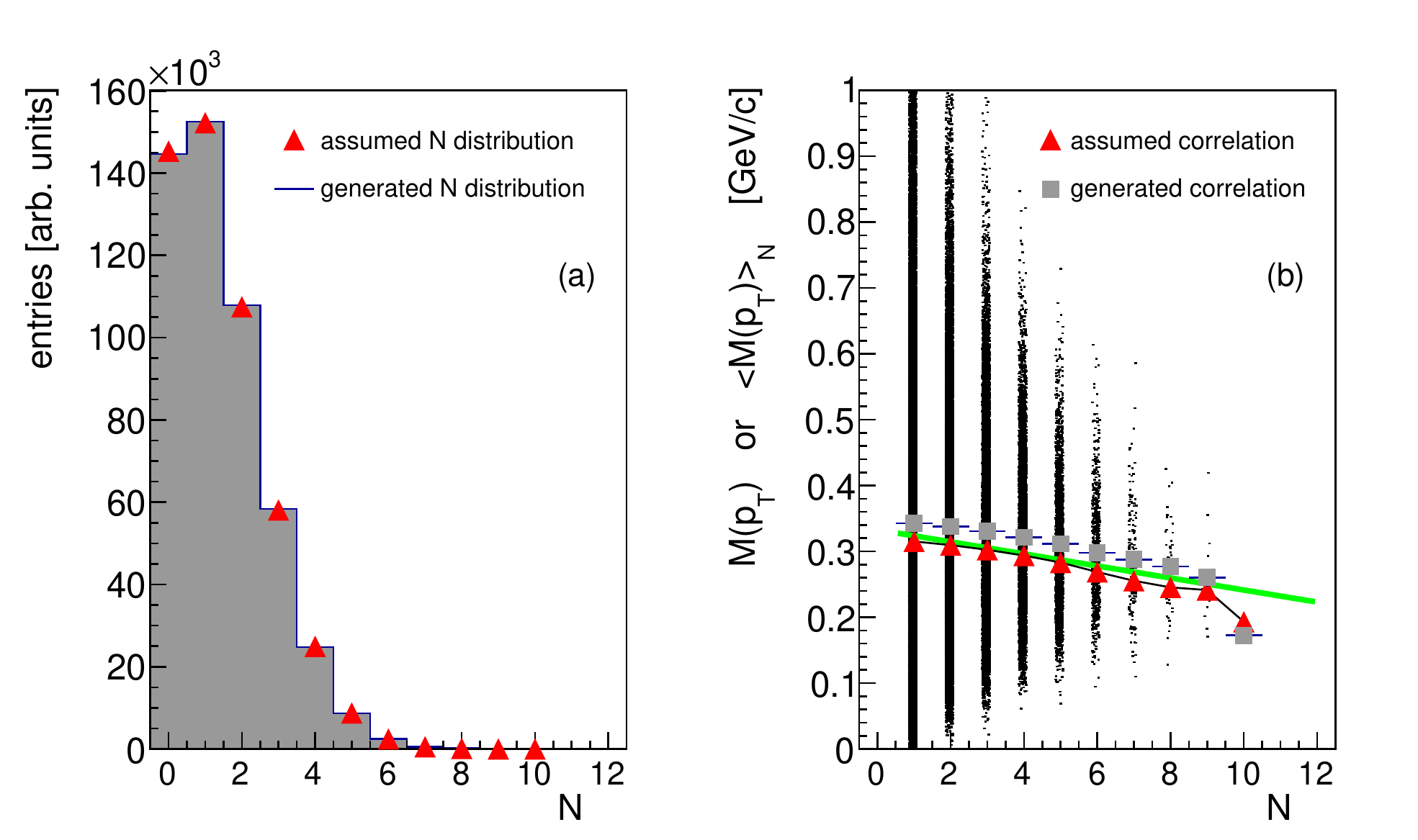}
\caption[]{(Color online) Properties of the fast generator producing $M(p_T)$
versus $N$ correlation (see the text for details). Green solid
line ({\it b}) shows Eqs. (\ref{T-N}, \ref{T-theta}), where
$\langle N \rangle$ = 1.4 and $T_N=\langle M(p_T) \rangle_{N}/2$.}
\label{figMptN}
\end{figure}

The results  of fast generators in the previous section showed the
same behavior and magnitudes of $\Delta[P_T, N]$ and $\Sigma[P_T,
N]$ measures.
The MC simulation, presented in this section, is introduced in
order to check whether one can propose a fast generator for which
different values of $\Delta[P_T, N]$ and $\Sigma[P_T, N]$ may be
obtained. As an example, we consider the $M(p_T)$ versus $N$
correlation suggested in Ref.~\cite{GM:1992}, where $M(p_T)$ is
the event mean single-particle transverse momentum and $N$ is the
particle  multiplicity.
In Fig.~\ref{figMptN} ({\it a}) the assumed multiplicity
distribution is presented as  red triangles (those values
correspond to the accepted multiplicities at forward rapidities in
$p+p$ collisions at the beam energy 158 GeV \cite{d1}). As seen,
the generated multiplicity distribution (gray histogram) coincides
with the assumed one. For each event, particle momenta are
generated from transverse momentum distribution (\ref{boltz}) with
$T$ taken as $T_N=\langle M(p_T) \rangle_{N}/2$, where $\langle
M(p_T) \rangle_{N}$ is dependent on generated multiplicity $N$ as
shown in Fig.~\ref{figMptN} ({\it b}) by the red triangles
\cite{d1}. The range of $p_T$ generation is from zero to 2
GeV/$c$. In Fig.~\ref{figMptN} ({\it b}) the scatter plot
represents all generated events ($M(p_T)$ values) and the gray
squares their profile histogram ($\langle M(p_T) \rangle_{N}$
values, where $\langle ... \rangle_{N}$ represents averaging
within the same multiplicity $N$).
The difference between red triangles (input values of $2\,T_N$ used in simulation)
and gray squares ($\langle M(p_T) \rangle _N$ values obtained from
simulated data set) is due to the fact that in transverse momentum
distribution used in simulation (\ref{boltz}) the average transverse momentum
is only approximately equal to $2\,T$. It was, however, verified
by an independent analysis that when using $f(p, T) = C\,p\,\exp (-p/T)$
distribution, for which the mean transverse momentum equals exactly
$2\,T$, red and gray points coincide. For the simulation presented
in Fig.~\ref{figMptN} the values of fluctuation measures obtained for
500 000 generated events are:
%
\eq{\label{DS-TN}
\Delta[P_T, N]~ =~ 0.8158~ \pm~ 0.0051~,~~~~\Sigma[P_T, N]~ =~
1.0075 ~\pm ~ 0.0018~.
}

\vspace{0.3cm}
Particle production considered in this section corresponds to
the distribution
\eq{\label{TNdistr}
F_N(p_1,\dots,p_N)~=~{\cal P}(N)\times f_N(p_1)\times \dots \times f_N(p_N)
}
where $f_N(p)=f(p,T_N)$ with $f(p,T_N)$ given by
Eq.~(\ref{boltz}), but with the parameter $T$ depending now on the
particle multiplicity, $T_N=T(N)$. The moments of single particle
distributions at fixed $N$ are then equal to
\eq{
  (\overline{p^k})_N~=~\int_0^{\infty} dp~p^k~f_N(p)
  ~.\label{pN}
}
The moments of single particle spectrum averaged over $N$ are
\eq{
\overline{p^k}~=~\sum_N \frac{N\,{\cal P}(N)}{\langle N\rangle}~ (\overline{ p^k})_N~,
}
and
$\omega[p]=(\overline{p^2}~-~\overline{p}^2)/\overline{p}$~.
With distribution (\ref{TNdistr}) one finds:
\eq{
& \langle P\rangle~=~\sum_N\,{\cal P}(N) \int f_N(p_1) dp_1\cdots \int f_N(p_N) dp_N~
(p_1+\cdots +p_N)~\nonumber \\
& =~\sum_N\, {\cal P}(N) \,N\, \int dp\,f_N(p)\,p~ =~\sum_N\,
{\cal P}(N) \,N\, (\overline{p})_N~ =~\overline{p}\cdot \langle
N\rangle~,
}
\eq{
 & \langle P^2\rangle~=~\sum_N\,{\cal P}(N) \int f_N(p_1)
dp_1\cdots \int f_N(p_N) dp_N~
(p_1+\cdots +p_N)^2~\nonumber \\
& =~\sum_N\, {\cal P}(N)\Big[ N\, \int dp\,f_N(p)\,p^2~+~N(N-1)\cdot
\Big(\int dp\,f_N(p)\,p\Big)^2\,\Big]~\nonumber \\
&=~\sum_N\, {\cal P}(N)\Big[ N\cdot (\overline{p^2})_N~+~N(N-1)\cdot
(\overline{p})_N^2\,\Big]~\nonumber \\
&=~\Big[\overline{p^2}-\overline{p}^2\Big]\cdot \langle N\rangle
~+~\sum_N{\cal P}(N)\,N^2\,(\overline{p})_N^2~,
}
\eq{
 & \langle P\,N \rangle~=~ \sum_N\,{\cal P}(N)\,N\, \int
f_N(p_1) dp_1\cdots \int f_N(p_N) dp_N~ (p_1+\cdots +p_N)
\nonumber \\
& =~\sum_N\, {\cal P}(N) \,N^2\, (\overline{p})_N~.
}
This gives:
\eq{
& \omega[P]~=~\omega[p]~+~
\frac{\sum_N{\cal P}(N)\,N^2\cdot (\overline{p})_N^2~-~
 \overline{p}^2\cdot \langle N\rangle^2~}{\overline{ p}\cdot \langle N\rangle}~,\\
& \langle PN\rangle~-~\langle P\rangle\,\langle N\rangle~=~
\sum_N\, {\cal P}(N)\,\cdot  N^2\cdot (\overline{p})_N ~-~\overline{p}\cdot \langle N\rangle^2~.
}
Finally,
\eq{
& \Delta[P,N]~=~1~+~\frac{\overline{ p}}{\omega[p]\cdot \langle N\rangle}~
\sum_N{\cal P}(N)\,N^2~\Big( Y_N^2~-~1\, \Big) ~,
\label{TN_corr_Delta}
\\
& \Sigma[P,N]~=~1~+~
\frac{\overline{ p}}{\omega[p]\cdot \langle N\rangle}~
\sum_N{\cal P}(N)\,N^2~\Big( Y_N~-~1\, \Big)^2 ~,
\label{TN_corr_Sigma}
 }
where
%
%
$Y_N \equiv (\overline{p})_N/\overline{p}$~.
%
%
Calculating numerically (\ref{TN_corr_Delta}) and
(\ref{TN_corr_Sigma}) with ${\cal P}(N)$ and  $T_N=\langle M(p_T) \rangle_{N}/2$
presented in Fig.~\ref{figMptN} ({\it a}) and ({\it b}),
respectively, one finds the $\Delta[P_T,N]$ and $\Sigma[P_T,N]$
values which coincide with those in Eq.~(\ref{DS-TN}) within
statistical uncertainties.

To make further analytical calculations several
simplifying assumptions will be adopted. First,
it will be assumed that produced particles
are massless. For $m=0$ in the distribution (\ref{boltz}) with
$T=T_N$ one finds
%
%
$(\overline{p})_N=2T_N~,~~ (\overline{p^2})_N=6T_N^2$~,
%
%
and
\eq{\label{p-av}
&\overline{p}~=~ \frac{1}{\langle N\rangle}\sum_N N\,{\cal
P}(N)\,(\overline{p})_N~=~2T\Big[1~+~\theta \Big(1~-~
 \frac{\langle N^2\rangle}{\langle N\rangle^2}\Big)\Big]~,
\\
& \overline{p^2}~=~\frac{1}{\langle N\rangle}\sum_N N\,{\cal
P}(N)\,(\overline{p^2})_N~=~ \frac{6T^2}{\langle N\rangle}\sum_N
N\,{\cal P}(N)\,\Big[1~+~\theta\cdot
\Big(1~-~\frac{N}{\langle N\rangle}\Big)\Big]^2\nonumber \\
&=~6T^2\Big[1~- ~\theta \cdot \frac{\langle N^2\rangle}{\langle
N\rangle^2}~+~ \theta^2\cdot\Big(1-2\frac{\langle N^2\rangle}
{\langle N\rangle^2}+ \frac{\langle N^3\rangle}{\langle
N\rangle^3}\Big)\Big]~.\label{p2-av}
}

Second, a
parametrization for the multiplicity dependent temperature
\eq{\label{T-N}
T_N~=~T\,\Big[1~+~\theta \cdot\Big(1~-~ \frac{N}{\langle
N\rangle}\Big)\Big]
}
proposed in Ref.~\cite{MRW:2004} will be adopted. This formula,
with small positive dimensionless parameter $\theta$, is
approximately valid for the data  in $p+p$ collisions at SPS
energy presented in Fig.~\ref{figMptN} ({\it b}).
Using the value of $\langle N\rangle =1.4$ (found from the data
in Fig.~\ref{figMptN} ({\it a})) the values of
\eq{\label{T-theta}
%
T~\cong~160~{\rm MeV}~, ~~~~\theta~\cong~ 0.04
}
are fixed from fitting the data  in
Fig.~\ref{figMptN} ({\it b}).
The correlation of the inverse slope ('temperature') parameter
$T_N$ versus $N$ in a form of Eq.~(\ref{T-N}) with $\theta >0$ is
probably of simple kinematic origin: when the multiplicity of
produced particles increases at fixed collision energy, there is
less and less energy to be transformed to transverse momenta of
produced particles. As a result, the average transverse momentum
per particle decreases when $N$ grows. However, in A+A collisions
the contribution of the transverse collective flow to particle
transverse momenta becomes important. This collective flow, in its
turn, increases with the number of produced particles. Therefore,
a correlation between $T_N$ and $N$ in a form (\ref{T-N}), but
with $\theta<0$, may be expected.

For further calculations we make the third simplification assuming
the Poisson shape for ${\cal P}(N)$ distribution. In this case one
obtains
\eq{\label{poisson}
&\langle N^2\rangle=\langle N\rangle^2 +\langle N\rangle~,~~~~
\langle N^3\rangle=\langle N\rangle^3 +3 \langle N\rangle^2+\langle N\rangle~,\\
&\langle N^4\rangle=\langle N\rangle^4 +6 \langle
N\rangle^3+7\langle N\rangle^2 +\langle N\rangle~,\label{poisson1}
}
and Eqs.~(\ref{p-av}) and (\ref{p2-av}) are transformed
to
\eq{\label{p-av-Poiss}
&\overline{p}~=~ 2T\Big[1~-~\frac{\theta}{\langle N\rangle}\Big]
~,~~~~
%
\overline{p^2}~=~6T^2\Big[1~-~\frac{2\theta}{\langle N\rangle}
~+~\theta^2 \cdot \Big(\frac{1}{\langle N\rangle}~+~
\frac{1}{\langle N\rangle^2}\Big)\Big] ~.
%
}
This gives
\eq{\label{omegap-TN}
&
\omega[p]~=~\frac{\overline{p^2}~-~\overline{p}^2}{\overline{p}}~\cong~
T\Big[1~-~\frac{\theta}{\langle N\rangle}\Big]~,~~~
%
%
Y_N~=~\frac{(\overline{p})_N}{\overline{p}}~\cong~ 1~+~\theta
\cdot \Big[1~-~\frac{N}{\langle N\rangle}\Big]~,
%
%
}
where the second and higher powers of $\theta$ have been neglected
and $\langle N\rangle\gg 1$ is assumed (this is our fourth and the
last simplification). For $\Sigma[P,N]$ (\ref{TN_corr_Sigma}) one
obtains
\eq{
& \Sigma[P,N]~=~1~+~ \frac{\overline{ p}}{\omega[p]\, \langle
N\rangle}~
\sum_N{\cal P}(N)\,N^2~\Big[Y_N~-~1 \Big]^2\nonumber \\
&  \cong~1~+~ 2\,\frac{\theta^2}{\langle N\rangle}\, \Big[ \langle
N^2\rangle-2\frac{\langle N^3\rangle}{\langle N\rangle}+
\frac{\langle N^4\rangle}{\langle
N\rangle^2}\Big]~=~1~+~2\,\theta^2~,
\label{S-m0}
}
where Eqs.~(\ref{poisson}) and (\ref{poisson1}) have been used at
the last step in Eq.~(\ref{S-m0}).
%
%
%
%

The $\Delta[P,N]$ (\ref{TN_corr_Delta}) is calculated as
\eq{\label{D-m0}
& \Delta[P,N]~=~1~+~
\frac{\overline{ p}}{\omega[p]\cdot \langle N\rangle}~
\sum_N{\cal P}(N)\,N^2~\Big[Y_N^2~-~1 \Big]\nonumber \\
&  \cong~1~+~ ~\frac{4\,\theta }{\langle N\rangle}\, \Big[
\Big[\langle N^2\rangle~-~\frac{\langle N^3\rangle}{\langle
N\rangle}\Big]~=~1-~ 4\,\theta~.
}

With $\theta=0.04$ (\ref{T-theta}) one obtains from
Eqs.~(\ref{S-m0}) and (\ref{D-m0}):
\eq{\label{DSm0TN}
\Sigma[P,N]~\cong~ 1.0032~,~~~~ \Delta[P,N]~\cong~ 0.8400~.
}
The results of our {\it approximate} analytical calculations (\ref{DSm0TN})
may be compared with the full MC calculations (\ref{DS-TN}).

Note that the correlation (\ref{T-N}) between $T_N$ and $N$ leads
to the additional term to $\Sigma$ (\ref{S-m0}) proportional to
$\theta^2$, whereas $\Delta$ (\ref{D-m0}) includes a linear
$\theta$-term. Therefore, the $\Delta[P,N]$ measure is much more
sensitive to the correlations (\ref{T-N}) between $T_N$ and $N$
than $\Sigma[P,N]$: the linear $\theta$ contribution is
essentially larger than $\theta^2$ one, as $\theta \ll 1$.
Besides, it is sensitive to a sign of $\theta$. Therefore, both
suppression (at $\theta>0$) and enhancement (at $\theta <0$)
effects for $\Delta[P,N]$ may be observed.

\section{Model Examples }\label{urqmd}

\subsection{Quantum Gases}
The  strongly intensive fluctuation measures  $\Delta[P_T,N]$ and
$\Sigma[P_T,N]$ have been recently studied in Ref.~\cite{GR:2013}
for the ideal Bose and Fermi gases within the grand canonical
ensemble. As it was already noted in Ref.~\cite{GGP:2013}, the
Boltzmann approximation satisfies the conditions of
the IPM, i.e. Eq.~(\ref{DelSig=1}) is valid. Quantum statistics introduces
particle correlations and the following general
relations have been found \cite{GR:2013}:
\eq{\label{BBF}
& \Delta^{{\rm Bose}}[P_T,N]~<~\Delta^{{\rm Boltz}}=1~<~\Delta^{{\rm Fermi}}[P_T,N]~,\\
& \Sigma^{{\rm Fermi}}[P_T,N]~<~\Sigma^{{\rm
Boltz}}=1~<~\Sigma^{{\rm Bose}}[P_T,N]~,\label{FBB}
}
i.e.  Bose statistics makes $\Delta[P_T,N]$ to be smaller and
$\Sigma[P_T,N]$ larger than unity, whereas Fermi statistics works
in exactly opposite way.
The Bose statistics of the pion gas appears to be the main source
of quantum statistics effects in the hadron gas with the
temperature  typical for the hadron system created in A+A
collisions. It gives approximately  $\Delta[P_T,N]\cong 0.8$ and
$\Sigma[P_T,N]\cong 1.1$, at $T\cong 150$~MeV, i.e.  suppression
of $\Delta[P_T,N]$ and enhancement of $\Sigma[P_T,N]$ in a
comparison to the Boltzmann approximation, equal to the IPM
results (\ref{DelSig=1}). Fermi statistics contributions to
$\Delta[P_T,N]$ and $\Sigma[P_T,N]$ for the protons are almost
negligible for typical temperatures and baryonic chemical
potentials in the hadron gas created in A+A collisions.

\subsection{UrQMD}

In this subsection we discuss the UrQMD \cite{urqmd} results.
In Ref.~\cite{GGP:2013} the simulations for
$\Delta[P_T, N_-]$ and $\Sigma[P_T, N_-]$,
where $N_-$ is the number of negative particles, were considered.
In the sample of 7\% most central Xe+La
collisions the fluctuation measure $\Sigma[P_T, N_-]$
appears to be close to 1 for the whole SPS
energy region $E_{lab}$ from 20 to 158 GeV per nucleon, whereas
the fluctuation measure $\Delta[P_T, N_-]$ increases
with the collision energy from the value of 1 at
$E_{lab}=20$~GeV per nucleon to approximately 1.4 at
$E_{lab}=158$~GeV per nucleon.
Note that the
UrQMD takes into account several sources of fluctuations and
correlations, e.g., exact conservation laws, resonance decays, flow effects, etc.

\begin{figure}[ht]
\centering
\includegraphics[width=0.49\textwidth]{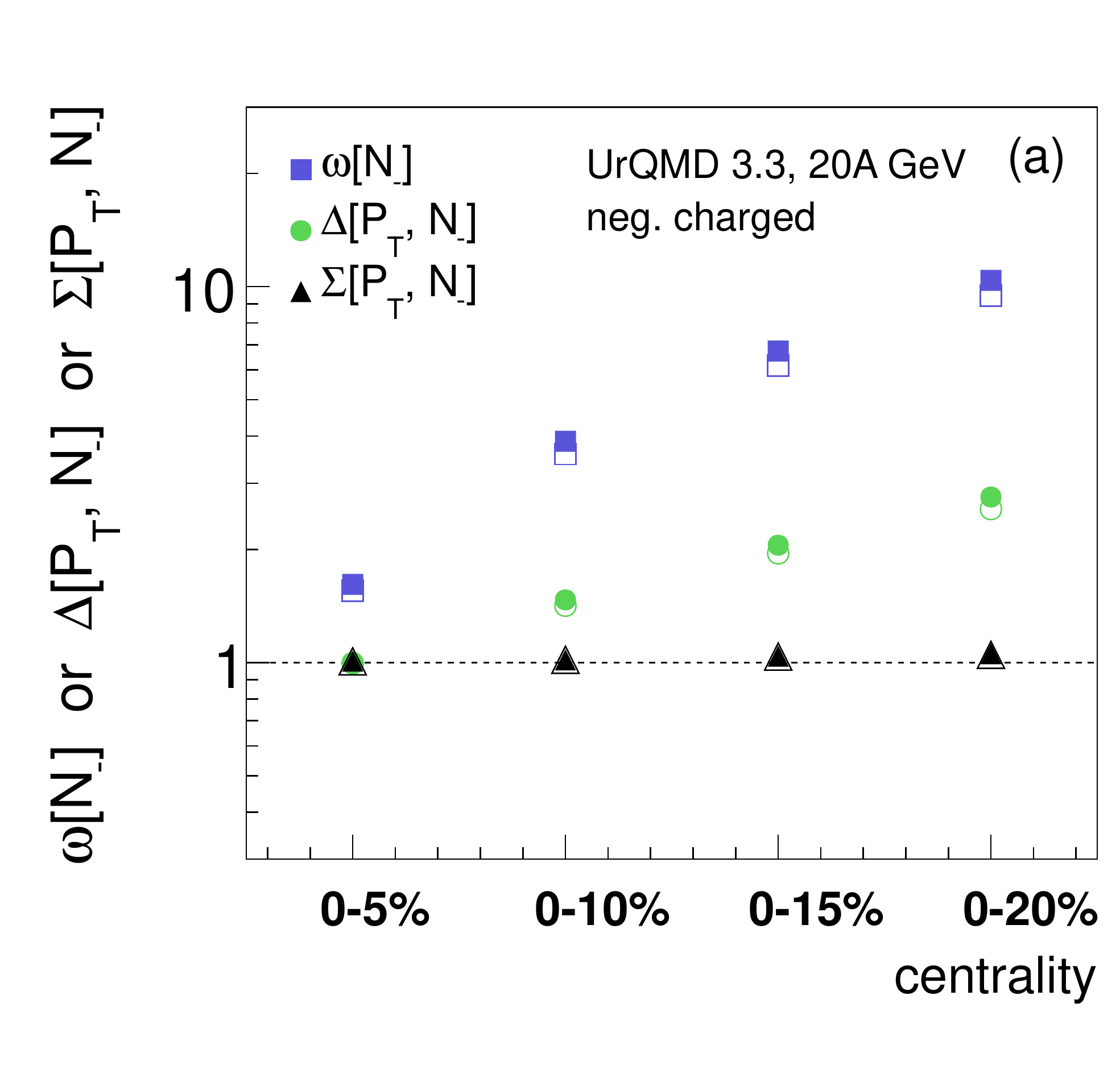}
\includegraphics[width=0.49\textwidth]{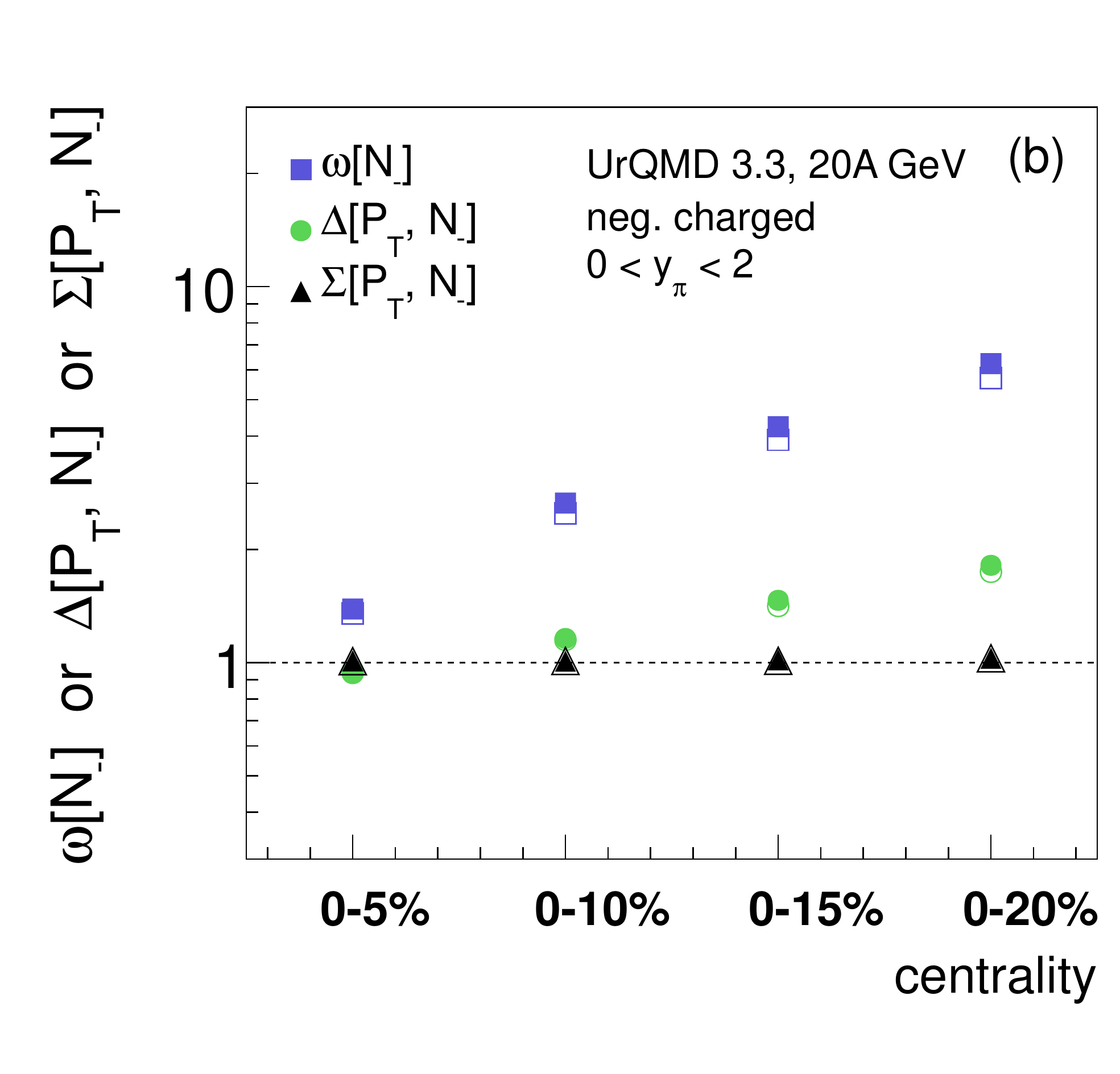}
\caption[]{(Color online) The UrQMD results for the
centrality dependence of $\omega[N_-]$ ({\it squares}),
$\Delta[P_T, N_-]$ ({\it circles}), and $\Sigma[P_T, N_-]$ ({\it triangles})
in Pb+Pb collisions at $E_{lab}=20$A GeV. A centrality selection
is done with a restriction on the impact parameter $b$. ({\it a}):
The full $4\pi$ detector acceptance. ({\it b}): Only particles
with center of mass rapidity in the interval $1<y_{\pi}<2$ are accepted (pion mass was assumed for all particles). Open symbols correspond to the case when 10\% of particles was randomly rejected.}
\label{figurqmd}
\end{figure}

We use the UrQMD simulations in Pb+Pb collisions at
$E_{lab}=20$~GeV per nucleon to study  $\Delta[P_T, N_-]$ and $\Sigma[P_T, N_-]$.
With this example we illustrate effects of the
centrality selection and limited detector acceptance and efficiency
in A+A collisions. The results presented in Fig.~\ref{figurqmd}
correspond to the centrality bins of 5\%, 10\%, 15\%, and 20\%  most
central Pb+Pb collisions. One observes very strong increase of $\omega[N_-]$
with a width of the centrality bin. This reflects the fact
that fluctuations of the number of nucleon participants affect strongly
the fluctuations of final hadron multiplicities. Therefore, scaled variances
as the fluctuation measures become almost useless for wide centrality bins.
For these wide samples of collisions, the scaled variances
do not describe physical properties of the system but
reflect the system size fluctuations
(see more details in Ref.~\cite{HSD}). The strongly intensive measures
$\Delta[P_T,N_-]$ and $\Sigma[P_T,N_-]$ look much more appropriate.
These quantities are not sensitive to the trivial system size fluctuations.
Their dependence on the size of the centrality bin  is rather moderate
(it is more pronounced for  $\Delta[P_T,N_-]$) and
reflects changes in local physical properties for different
centrality samples.

Another important aspect of today measurements of the e-by-e fluctuations
in A+A collisions is a limited detector acceptance and/or efficiency. Particles may be lost due to the geometry of the detector (for example fixed target experiments typically cover only forward hemisphere) and we call it {\it acceptance losses}. On the other hand, even in this accepted kinematic region we still may have {\it efficiency losses} due to track reconstruction problems (including problems with ionization energy loss, $dE/dx$, reconstruction).

The UrQMD results for negatively charged particles in Pb+Pb
collisions at $E_{lab}=20$A~GeV for the full 4$\pi$ acceptance and for
the particles accepted only in the center of mass rapidity interval $1<y_{\pi}<2$
are shown in Fig.~\ref{figurqmd} ({\it a}) and ({\it b}), respectively (full symbols).
From a comparison of the results for the full and limited detector acceptance
one observes rather strong effects of acceptance losses for the
scaled variance $\omega[N_-]$. The strongly intensive measures
$\Delta[P_T,N_-]$ and $\Sigma[P_T,N_-]$ look again more appropriate.
The effects of the limited acceptance are rather moderate
for $\Delta[P_T,N_-]$ and almost absent for $\Sigma[P_T,N_-]$.
We also would like to stress that the acceptance dependence shown in Fig.~\ref{figurqmd} is the example only. In general, the measured magnitude of $\omega[N]$, $\Delta[P_T,N]$ or $\Sigma[P_T,N]$ depends on both the correlation(s) length(s) and the size of the acceptance region (when the kinematic acceptance is much smaller that the correlation range the effect will be washed out). Therefore, when comparing experimental results to models the experimental kinematic restrictions should be carefully taken into account.

Finally, the example of the effect of efficiency losses is shown by open symbols in Fig.~\ref{figurqmd}. In this case from each event we randomly rejected 10\% of particles. As seen, the effect of efficiency losses is small or even negligible (comparison of full and open symbols) for all presented fluctuation measures but, in general, it depends on the fraction of rejected particles.

\section{Summary}\label{sum}
In the present paper  strongly intensive measures
of the event-by-event fluctuations $\Delta[P_T,N]$ and
$\Sigma[P_T,N]$ are studied.
The recently proposed special normalization for these fluctuation
measures are used, and  it ensures that these measures are
dimensionless and yields a common scale required for a
quantitative comparison of fluctuations.
%
Several phenomenological models are considered using the Monte
Carlo simulations and analytical calculations. Our studies include
different versions of the model of independent sources: with fixed
number of sources, with the Poisson distribution of the number of
sources, and with the Negative Binomial distribution.
The
quantities $\Delta[P_T,N]$ and $\Sigma[P_T,N]$ are found to be
independent of the average number of sources and of its
fluctuations. This reflects the strongly intensive
properties of the $\Delta$ and $\Sigma$ measures, and is a
main motivation of their using for the analysis of the
event-by-event fluctuations in nucleus-nucleus collisions.
The transverse momentum distribution of particles emitted from the
source are assumed to be a thermal-like (Boltzmann) distribution
over transverse mass. The average single-particle transverse
momentum is then controlled by the inverse slope (temperature)
parameter.

The system of sources with constant temperature appear to be
equivalent to the model of independent sources, i.e. a relation
$\Delta[P_T,N]=\Sigma[P_T,N]=1$ is obtained. For independent
temperature fluctuations from source to source, one finds the
correlations between transverse momenta of particles emitted from
the same source. This leads to
$\Delta[P_T,N]=\Sigma[P_T,N]=1+q_S$, where the value of $q_S$ is
positive and depends only on the parameters of a single source.
If all sources have the same fluctuating temperature,
the model of independent sources becomes no more applicable. One
obtains $\Delta[P_T,N]=\Sigma[P_T,N]=1+Q_S$, where the value of
$Q_S$ increases linearly with the average number of sources
$\langle N_S\rangle$ and increases with $\sigma_T$ which determines
the size of temperature fluctuations.

A model which introduces a correlation between the temperature
parameter and particle multiplicity is studied. In this case, the
different values for the $\Delta$ and $\Sigma$ measures have been
found: $\Delta[P_T,N]=1+q_{\delta}$ and
$\Sigma[P_T,N]=1+q_\sigma$.
Analytical calculations under several simplifying assumptions
give: $q_\delta\cong -\, 4\theta$ and $q_\sigma\cong 2\theta^2$~,
where the parameter $\theta$ describes the correlations between $T_N$
and $N$ according to (\ref{T-N}) and is assumed to be small,
$|\theta| \ll 1$.
%

The UrQMD simulations for Pb+Pb collisions at the collision energy
$E_{lab}=20$~GeV per nucleon are done and analyzed. With this example we illustrate
a role of the centrality selection and limited detector acceptance and efficiency
in A+A collisions. We find that the strongly intensive quantities
$\Delta[P_T,N]$ and $\Sigma[P_T,N]$ have an advantage over the standard
fluctuation measures. In contrast to the scaled variance, $\Delta[P_T,N]$ and $\Sigma[P_T,N]$
demonstrate much weaker sensitivity to the width of the centrality bin
and to the limited detector acceptance and efficiency.

In all considered model examples, $\Delta[P_T,N]$
appears to be more sensitive to interparticle correlations
than $\Sigma[P_T,N]$. This reveals itself as stronger deviations
of $\Delta[P_T,N]$ from the IPM results (\ref{DelSig=1}).
Even for $\Delta[P_T,N]=\Sigma[P_T,N]$, in the MC simulations
in Sec.~\ref{fast}, a stronger sensitivity of $\Delta[P_T,N]$
manifests as its larger statistical errors.

We hope that the results obtained in this paper will be helpful to
elucidate the properties of $\Delta[P_T,N]$ and $\Sigma[P_T,N]$
measures.

\begin{acknowledgments}
We would like to thank Marek Ga\'zdzicki, Laszlo Jenkovszky, and
Stanislaw Mr\'owczy\'nski for fruitful discussions and comments.
We are indebted to the authors of the UrQMD model for the possibility
to use their code in our analysis.
The work of M.I.G. is supported by the
State Agency of Science, Innovations and
Informatization of Ukraine, contract F58/384-2013.
The work of K.G. was supported by the the National Science Center,
Poland grant DEC-2011/03/B/ST2/02617 and grant 2012/04/M/ST2/00816.
\end{acknowledgments}



\end{document}